\documentclass[11pt]{article}%
\usepackage{graphicx}
\usepackage[misc,geometry]{ifsym}
\usepackage{mathptmx}       
\usepackage{amsfonts,amsmath,array,amssymb,latexsym}
\newtheorem{t1}{Theorem}[section]
\newtheorem{p1}{Proposition}[section]

\newtheorem{c1}{Corollary}[section]
\newtheorem{d1}{Definition}[section]
\newtheorem{r1}{Remark}[section]

\newtheorem{ex}{Example}[section]

\title{\huge\bf
On cumulative residual (past) inaccuracy for truncated random variables\thanks{
The work of C.\ Kundu is supported by Department of Science and Technology, Government of India 
(Ref.\ No.\ SR/FTP/MS-016/2012) and the research by A.\ Di Crescenzo and M.\ Longobardi is partially 
supported by GNCS-INdAM and Regione Campania (Legge 5).}
\date{
Author's version.  Published in:  {\em Metrika}\ 79 (2016), pp.\ 335-356,   
doi:   10.1007/s00184-015-0557-5 \ -- \ 
URL: https://link.springer.com/content/pdf/10.1007/s00184-015-0557-5.pdf}
}

\author{
\large \bf Chanchal Kundu\footnote{ 
Department of Mathematics, Rajiv Gandhi Institute of Petroleum Technology, 
Rae Bareli 229 316, U.P., India, 
E-mail: ckundu@rgipt.ac.in; chanchal$_{-}$kundu@yahoo.com }
\qquad
Antonio Di Crescenzo\footnote{  
Dipartimento di Matematica, Universit\`a di Salerno,
Via Giovanni Paolo II, n. 132, I-84084 Fisciano (SA), Italy }
\qquad
Maria Longobardi\footnote{  
Dipartimento di Matematica e Applicazioni, Universit\`a di Napoli Federico II,
Via Cintia, I-80126 Napoli, Italy }
}

\begin{document}

\maketitle

\begin{abstract}
To overcome the drawbacks of Shannon's entropy, the concept of cumulative residual and past entropy
has been proposed in the information theoretic literature. Furthermore, the Shannon entropy has
been generalized in a number of different ways by many researchers. One important extension is Kerridge
inaccuracy measure. In the present communication we study the cumulative residual and past
inaccuracy measures, which are extensions of the corresponding cumulative entropies.
Several properties, including monotonicity and bounds, are obtained for left, right and doubly
truncated random variables.


\end{abstract}

\section{Introduction and preliminary results}
The concept of using the cumulative distribution function of a random variable to define its information
content was first introduced by Rao et al.\ (2004). In recent years, there has been a great interest in the
measurement of uncertainty of probability distributions. It is well-known that the traditional measure of
uncertainty contained in a random variable $X$ is the Shannon's (1948) differential entropy which has
mushroomed into a large body of knowledge revolutionizing many areas such as financial analysis,
data compression, statistics, and information theory.
\par
\hspace*{.2in}
Let $X$ be an absolutely continuous nonnegative random variable with support $(0,\infty)$,
probability density function $f$, distribution function $F(x)$ and reliability function $\overline F(x)=1-F(x)$.
Then the Shannon entropy (also known as differential entropy) is defined as
\begin{eqnarray}\label{eq1}
H(X)=-\int_{0}^\infty f(x)\ln f(x) dx.
\end{eqnarray}
In spite of its enormous success, this measure has some drawbacks and in certain situations it may
not be appropriate. For example, Shannon entropy (\ref{eq1}) may take any value on the extended
real line and is defined only for distributions possessing a density function (see Rao et al., 2004,
for other details). To get rid of these drawbacks an alternative measure of uncertainty, called
{\em cumulative residual entropy} (CRE), has been proposed by Rao et al.\ (2004) as follows:
\begin{eqnarray}\label{eq2}
\varepsilon(X)=-\int_{0}^{\infty}\overline F(x)\ln{\overline F(x)}dx.
\end{eqnarray}
\hspace*{.2in}
This measure is defined similarly as the Shannon's entropy for lifetime distributions, in the sense that
it takes into account the reliability function $\overline F(x)$ instead of  the density function $f(x)$.
In this case the measurement of uncertainty is based on cumulative information rather than local
information. Some properties and applications of CRE in reliability engineering and computer
vision have been also studied by Rao et al.\ (2004) and Rao (2005).
\par
\hspace*{.2in}
We recall that if $X$ is a  random variable with support $(0,\infty)$ and finite expectation $E(X)$,
then the equilibrium random variable of $X$ is usually denoted by $X_e$, and has density
\begin{equation}
 f_e(x)= \frac{\overline F(x)}{E(X)}, \qquad x\in (0,\infty)
\label{eq:densfequil}
\end{equation}
(see Gupta, 2007, and references therein, for instance).
The  equilibrium distribution arises as the limiting distribution of the forward recurrence time in
renewal processes, and thus it deserves interest in various applications in reliability and queueing.
Hereafter we pinpoint the connection between the CRE and the entropy of the equilibrium distribution.
\begin{p1}\label{prop:p1}
If $X$ is a nonnegative random variable having support $(0,\infty)$ and finite expectation $E(X)$, then
the following identity holds:
\begin{equation}
\varepsilon(X)= E(X) \{H(X_e)-\ln E(X)\},
\label{eq:varXHXe}
\end{equation}
where $H(X_e)$ is the Shannon entropy of the equilibrium distribution of $X$.
\end{p1}
Proof: The proof follows from identity
$H(X_e)=- \int_{0}^\infty f_e(x)\ln f_e(x) dx$, with $f_e(x)$ given in (\ref{eq:densfequil}),
after straightforward calculations.
$\hfill\square$
\par
\hspace*{.2in}
Clearly, from (\ref{eq:varXHXe}) we have that the CRE is a linear increasing transformation of the
Shannon entropy of the equilibrium distribution. Specifically, from Proposition \ref{prop:p1} we see
that $\varepsilon(X)$  is, apart from a constant term, a measure of the entropy of $X_e$ in the unity
measure of $E(X)$. Indeed, if $E(X)=1$ then $\varepsilon(X)=H(X_e).$
\par
\hspace*{.2in}
Recently, Di Crescenzo and Longobardi (2009) introduced an information measure based
on the cumulative distribution function $F(x)$, called {\em cumulative past entropy} (CPE) and defined as:
\begin{eqnarray}\label{eq3}
\overline\varepsilon(X)=-\int_{0}^{\infty}F(x)\ln{F(x)}dx.
\end{eqnarray}
\hspace*{.2in}
Furthermore, numerous definitions and generalizations of (\ref{eq1}) have been proposed in the
literature. An important development in this direction is the {\em inaccuracy measure} due to
Kerridge (1961), which involves two absolutely continuous nonnegative random variables
$X$ and $Y$ with support $(0,\infty)$, and having distribution functions $F(x),~G(x)$,
reliability functions $\overline F(x),~\overline G(x)$ and probability density functions
$f,~g$, respectively. If $f(x)$ is the actual density corresponding to the
observations and $g(x)$ is the density assigned by the experimenter, then the
inaccuracy measure of $X$ and $Y$ is given by
\begin{eqnarray}\label{eq4}
 H_{X,Y}=-\int_{0}^\infty f(x)\ln g(x) dx.
\end{eqnarray}
It has applications in statistical inference, estimation and coding theory. Clearly, if $g(x)=f(x)$
then (\ref{eq4}) reduces to (\ref{eq1}).
\par
\hspace*{.2in}
Analogous to CRE and CPE the following information measures can be considered.
Let $X$ and $Y$ be nonnegative random variables having support $(0,\infty)$, distribution
functions $F(x)$ and $G(x)$, reliability functions $\overline F(x)$ and $\overline G(x)$,
respectively. Then, the {\em cumulative residual inaccuracy} (CRI) is defined as
\begin{eqnarray}\label{eq5}
\mathcal{C}H_{X,Y}=-\int_{0}^\infty \overline F(x)\ln\overline G(x) dx;
\end{eqnarray}
the {\em cumulative past inaccuracy} (CPI) is defined as
\begin{eqnarray}\label{eq6}
\mathcal{C}\overline H_{X,Y}=-\int_{0}^\infty F(x)\ln G(x) dx.
\end{eqnarray}
\par\hspace*{.2in}
Similarly as in (\ref{eq2}) and (\ref{eq3}), the basic idea is to replace the density function by survival
(distribution) function in Kerridge inaccuracy measure. Also, the measures given in (\ref{eq5}) and
(\ref{eq6}) are defined even if $X$ and $Y$ do not possess a probability density. Moreover, in many
practical situations the distribution function deserves larger interest and is observable.
For example, if the random variable is the life span of a machine, then the event of main interest
is whether the life span exceeds $t$, rather than it equals $t$. It is to be noted that (\ref{eq5}) and (\ref{eq6})
can be viewed as the cumulative analogue of Kerridge inaccuracy measure and represent the information content when
using $G(x)$, the distribution asserted by the experimenter due to missing/incorrect information in expressing
statement about probabilities of various events in an experiments, instead of true distribution $F(x)$.
\par\hspace*{.2in}
In analogy with Proposition \ref{prop:p1} we are now able to state the following result, which
relates the CRI to the inaccuracy measure of the equilibrium distributions. The proof is
omitted being similar.
\begin{p1}\label{prop:2chxy}
Let $X$ and $Y$ be nonnegative random variables having support $(0,\infty)$ and finite
expectations $E(X)$ and $E(Y).$
Let $f_e(x)=\frac{\overline F(x)}{E(X)}$, $x>0$,  and $g_e(x)=\frac{\overline G(x)}{E(Y)}$, $x>0$,
be the densities of the equilibrium distributions of  $X$ and $Y$, respectively. Then,
\begin{equation}
 \mathcal{C}H_{X,Y}= E(X) \{H_{X_e,Y_e}-\log E(Y)\},
\end{equation}
where $H_{X_e,Y_e}=- \int_{0}^\infty f_e(x)\ln g_e(x) dx.$
\end{p1}
\par\hspace*{.2in}
Propositions \ref{prop:2chxy} shows that $\mathcal{C}H_{X,Y}$ expresses, apart from
a constant term,  the inaccuracy measure of $X_e$ and $Y_e$  in the unity
measure of $E(X)$. Indeed, if $E(X)=1$ and $E(Y)=1$ then $\mathcal{C}H_{X,Y}= H_{X_e,Y_e}$.
\par\hspace*{.2in}
We also recall the Kullback-Leibler distance of $X$ and $Y$, defined as
$$
 K\! L(X,Y):=   H(X)-H_{X,Y}= -\int_{0}^\infty f(x)\ln \frac{f(x)}{g(x)} dx.
$$
This is another  quantity of interest in information theory, which can be viewed as
the ``information'' lost when the density $g$ is used to approximate $f$.
Let us now express the Kullback-Leibler distance of equilibrium distributions in terms
of CRE and CRI. (The proof is omitted for brevity.)
\begin{p1}
Let $X$ and $Y$ be nonnegative random variables having support $(0,\infty)$ and expectations
$E(X)$ and $E(Y).$  Then,
\begin{equation}\label{eq:KLe}
 K\! L(X_e,Y_e)= \log \frac{E(X)}{E(Y)} +\frac{1}{E(X)} \{\varepsilon(X) -\mathcal{C}H_{X,Y}\}.
\end{equation}
\end{p1}
\par\hspace*{.2in}
Hence, we note that if $E(X)=1$ and $E(Y)=1$ then $K\! L(X_e,Y_e)= \varepsilon(X) -\mathcal{C}H_{X,Y}$.
\par\hspace*{.2in}
The following example  illustrates the role of CRI and CPI in the comparison of random lifetimes
having exponential and Erlang(2) distributions. In particular, it is shown an instance in which
$H_{X,Y}=H_{Y,X}$ even if the measures defined in (\ref{eq5}) and (\ref{eq6}) take different values
when the role of $X$ and $Y$ is interchanged.
\begin{ex}
Let $X$ and $Y$ denote random lifetimes of two components with probability density functions
$f(x)=e^{-x},~x\in(0,\infty)$ and $g(x)=\lambda^2 xe^{-\lambda x},~x\in(0,\infty)$, $\lambda>0$,
respectively. By simple calculations, from (\ref{eq4}) we have
$H_{X,Y}=\gamma+\lambda -2\ln \lambda$, where $\gamma\simeq 0.577216$ is the Euler's
constant, and $H_{Y,X}=2/\lambda$. Let $\lambda$ be the solution of the transcendental equation
$\gamma+\lambda -2\ln \lambda-2/\lambda=0$, i.e.\ $\lambda\simeq 0.624182$. Hence, in this instance
we have $H_{X,Y}=H_{Y,X}$, so that the Kerridge inaccuracy measure doesn't bring out any differences
between these two cases. However, from (\ref{eq5}) we have $\mathcal{C}H_{X,Y}=0.809178$
and $\mathcal{C}H_{Y,X}=1.13724$. Therefore, the inaccuracy measure of the observer for the
observations $X$ (resp. $Y$) taking $Y$ (resp. $X$) as corresponding assigned outcomes by the
experimenter are identical. Nevertheless, $\mathcal{C}H_{X,Y}<\mathcal{C}H_{Y,X}$, i.e., the CRI
of the observer for $X,~Y$ is lower than that for $Y,~X$. Similarly, their CPIs are also different; indeed
from (\ref{eq6}) we have $\mathcal{C} \overline H_{X,Y}=0.955988$ and $\mathcal{C} \overline H_{Y,X}=0.458129$.
$\hfill\square$
\end{ex}
\hspace*{.2in}
We recall that for a nonnegative random variable $X$ with support $(0,\infty)$, the cumulative
hazard rate and the cumulative reversed hazard rate are defined respectively as
$$
 R_F(x)=-\ln  \overline F(x)=\int_0^x \lambda_F(t)dt,
 \qquad
 T_F(x)=-\ln F(x)=\int_x^{\infty} \phi_F(t)dt,
 \qquad x>0,
$$
where $\lambda_F(t)=f(t)/\overline F(t)$ is the hazard rate function of $X$, and $\phi_F(t)=f(t)/F(t)$
is the reversed hazard rate function of $X$. Let $R_G(x)$ and $T_G(x)$ be similarly defined for $Y$.
In order to pinpoint a probabilistic meaning of CRI and CPI let us now consider the following functions,
defined for $x>0$:
\begin{equation}
 R^{(2)}_F(x)=\int_0^x R_F(t)dt
 =-\int_0^x \ln \overline F(t) dt,
 \quad
 R^{(2)}_G(x)=\int_0^x R_G(t)dt
 =-\int_0^x \ln \overline G(t) dt,
 \label{eq:defR2}
\end{equation}
\begin{equation}
 T^{(2)}_F(x)=\int_x^{\infty} T_F(t)dt
 =-\int_x^{\infty}   \ln F(t) dt,
 \quad
 T^{(2)}_G(x)=\int_x^{\infty} T_G(t)dt
 =-\int_x^{\infty} \ln  G(t) dt.
 \label{eq:defT2}
\end{equation}
We thus note that the functions introduced in (\ref{eq:defR2}) and (\ref{eq:defT2}) are related to
quantities of interest in reliability theory
(see Barlow and Proschan, 1975, and Shaked and Shanthikumar,  2007, for details). We are now
able to express $\mathcal{C}H_{X,Y}$ and $\mathcal{C}\overline H_{X,Y}$ as suitable expectations.
\begin{p1}\label{prop:p4}
Let $X$ and $Y$ be nonnegative random variables having support $(0,\infty)$. Then,
\begin{equation}
 \mathcal{C}H_{X,Y}=E\left[R^{(2)}_G(X)\right],
 \qquad
 \mathcal{C}\overline H_{X,Y}=E\left[T^{(2)}_G(X)\right].
 \label{eq:espr}
\end{equation}
\end{p1}
Proof: Recalling (\ref{eq5}) and (\ref{eq6}), the proof of identities (\ref{eq:espr}) follows from
Eqs.\ (\ref{eq:defR2}) and (\ref{eq:defT2}) after straightforward calculations, similarly as
Proposition 2.1 of Di Crescenzo and Longobardi (2013).
$\hfill\square$
\par
\par
\hspace*{.2in}
The considered measures $\mathcal{C}H_{X,Y}$ and $\mathcal{C}\overline H_{X,Y}$ are useful
for comparing: \\
{\em (i)} \ the true density $f$ to the used density $g$ in statistical modeling, \\
{\em (ii)} \ the lifetime distributions of two independent components in reliability modeling. \\
In case {\em (i)} only  $\mathcal{C}H_{X,Y}$ and $\mathcal{C}\overline H_{X,Y}$ are meaningful.
In such a case the role of $\mathcal{C}H_{X,Y}$ emerges from Proposition \ref{prop:2chxy}, whereas
the meaning of $\mathcal{C}\overline H_{X,Y}$ can be similarly obtained on the ground of analogous
results provided in Park et al. (2012) and in Di Crescenzo and Longobardi (2015).
In case {\em (ii)} in addition to $\mathcal{C}H_{X,Y}$ and $\mathcal{C}\overline H_{X,Y}$
it is also useful to consider $\mathcal{C}H_{Y,X}$ and $\mathcal{C}\overline H_{Y,X}$, since
these measures are not symmetric.
Namely, $\mathcal{C}H_{X,Y}$ measures an information amount carried when $F$ is the true
distribution and is compared with $G$, whereas their role is inverted for $\mathcal{C}H_{Y,X}$;
a similar remark holds for $\mathcal{C}\overline H_{X,Y}$.
This is also confirmed by the results given in Proposition \ref{prop:p4}.
For instance, condition $\mathcal{C}H_{X,Y}<\mathcal{C}H_{Y,X}$ means that
$E\left[R^{(2)}_G(X)\right]<E\left[R^{(2)}_F(Y)\right]$, and thus the information amount
carried by $X$ with respect to $Y$ is smaller than that carried by $Y$ with respect to $X$.
In agreement with analogous measures, the use of $\mathcal{C}H_{X,Y}$ is suggested  when
$F$ is the actual distribution corresponding to the observations and $G$ is the distribution
chosen by the experimenter.

\par\hspace*{.2in}
The functions defined in (\ref{eq:defR2}) and (\ref{eq:defT2}) can also be used to express CRE and
CPE as means. Indeed, from (\ref{eq2}) and (\ref{eq3}) we have
\begin{equation}
 \varepsilon(X)=E\left[R^{(2)}_F(X)\right],
 \qquad
 \overline\varepsilon(X)=E\left[T^{(2)}_F(X)\right],
 \label{eq:entrop}
\end{equation}
in agreement with Proposition 3.1 of Di Crescenzo and Longobardi (2009).
The equalities shown in Eqs.\ (\ref{eq:espr}) and (\ref{eq:entrop}) suggest to introduce
the following suitable ratios.
\begin{d1}\label{def2}
Let $X$ and $Y$ be nonnegative random variables having support $(0,\infty)$.
Then, the {\em cumulative residual inaccuracy ratio} (CRIR) is defined as
\begin{eqnarray}\label{eqR1}
 {\cal CR}_{X,Y}
 =\frac{\mathcal{C}H_{X,Y}}{\varepsilon(X)}
 =\frac{E\left[R^{(2)}_G(X)\right]}{E\left[R^{(2)}_F(X)\right]};
\end{eqnarray}
the {\em cumulative past inaccuracy ratio} (CPIR) is defined as
\begin{eqnarray}\label{eqR2}
  {\cal C\overline R}_{X,Y}
  =\frac{\mathcal{C}\overline H_{X,Y}}{\overline\varepsilon(X)}
  =\frac{E\left[T^{(2)}_G(X)\right]}{E\left[T^{(2)}_F(X)\right]}.
\end{eqnarray}
\end{d1}
\par
\hspace*{.2in}
The above ratios give adimensional measures of closeness between $X$ and $Y$. Clearly,
we  have  ${\cal CR}_{X,Y}={\cal C\overline R}_{X,Y}=1$ if $X$ and $Y$ are identically distributed.
Moreover, recalling that the Kullback-Leibler distance is nonnegative, from (\ref{eq:KLe}) we
obtain the following upper bound:
$$
 {\cal CR}_{X,Y}\leq 1+\frac{E(X)}{\varepsilon(X)}\ln \frac{E(X)}{E(Y)}.
$$
Similar results can be obtained by resorting to the extensions of Kullback-Leibler information
investigated in Di Crescenzo and Longobardi (2015).
In the following example  the measures defined in (\ref{eqR1}) and (\ref{eqR2}) are employed to compare
suitable lifetime distributions.
\begin{ex}\label{ex1}
Let $X$ be exponentially distributed with mean 1, and $Y$ have {\em (i)} Weibull density $g(x)=r x^{r-1}e^{-x^r}$,
$x\in(0,\infty)$, and {\em (ii)} gamma density $g(x)=\frac{1}{\Gamma(r)}x^{r-1}e^{-x}$, $x\in(0,\infty)$,
where in both cases $Y$ has scale 1 and shape $r>0$. Figure \ref{fig1} shows the cumulative residual
and past inaccuracy ratios for $(X,Y)$ and $(Y,X)$. We note that such measures are not monotonic in $r$.
%
\begin{figure}[t]
\centering
\includegraphics[width=5.4 cm,keepaspectratio]{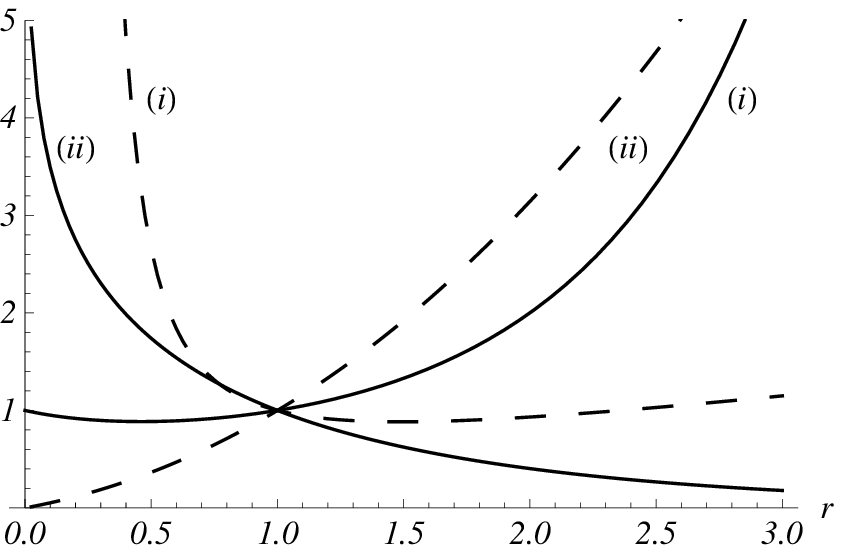}
$\,$
\includegraphics[width=5.4 cm,keepaspectratio]{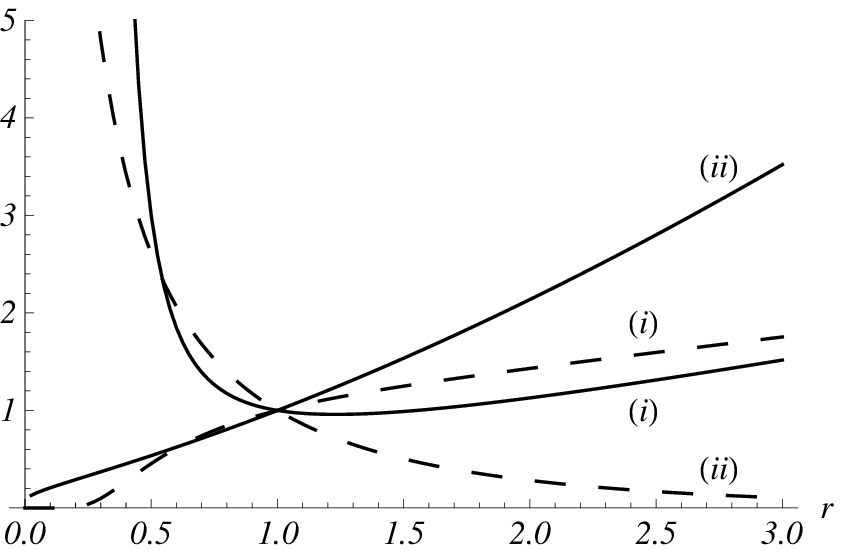}
\vspace{-.1cm}
\caption{Plots of CRIR and CPIR when $X$ has exponential density with mean 1 and $Y$ has
{\em (i)} Weibull density with parameters $(1,r)$, and {\em (ii)} gamma density with parameters
$(1,r)$, for $r\in(0,3)$ (cf.\ Example \ref{ex1}).
Left picture: ${\cal CR}_{X,Y}$ (full line) and ${\cal C\overline R}_{X,Y}$ (dashed line).
Right picture: ${\cal CR}_{Y,X}$ (full line) and ${\cal C\overline R}_{Y,X}$ (dashed line).
}\label{fig1}
\end{figure}
\end{ex}
\hspace*{.2in}
We remark that ${\cal CR}_{X,Y}$ and ${\cal C\overline R}_{X,Y}$ are not symmetric and thus, for instance,
${\cal CR}_{X,Y}$ and ${\cal CR}_{Y,X}$ have a different meaning. Roughly speaking, ${\cal CR}_{X,Y}$
measures the discrepancy in the information amount carried by the cumulative residual entropy when
the true distribution $F$ is replaced by a different distribution $G$.
Finally, in brief we note that ${\cal CR}_{X,Y}<1$ means that using the distribution $G$ instead of $F$ gives
less information in the sense of CRI rather than that carried by CRE of $F$. A similar remark can be
given for ${\cal C\overline R}_{X,Y}$.
\par
\hspace*{.2in}
In several contexts related to reliability theory dynamical measures are useful to describe the information
content carried by random lifetimes as age varies. This led several authors to deal with dynamic
information measures. See, for instance, Asadi and Zohrevand (2007), Chamany and Baratpour (2014),
Di Crescenzo and Longobardi (2009), Kundu and Nanda (2014), Misagh and Yari (2011),
Navarro et al.\ (2010), Sunoj et al.\ (2009). Dynamic versions of CRE and CPE have also
been proposed in the literature. Hereafter we consider CRI and CPI for truncated random variables.
\par
\hspace*{.2in}
The rest of the paper is arranged as follows. In Section 2 we study some properties of CRI and
dynamic CRI. Some bounds and inequalities are obtained. Analogous discussion is made for CPI
and dynamic CPI in Section 3. Section 4 is devoted to the study of CRI and CPI for doubly truncated
random variables. Conclusions are finally presented in Section 5.
\section{Results on (dynamic) CRI}
Asadi and Zohrevand (2007) considered the dynamic version of CRE, called {\em dynamic cumulative
residual entropy} (DCRE), which is defined as CRE of the residual lifetime $[X-t|X>t]$,  i.e.
\begin{eqnarray}\label{eq2.1}
\varepsilon(X;t)=-\int_{t}^{\infty}\frac{\overline F(x)}{\overline F(t)}\ln\frac{\overline F(x)}{\overline F(t)}dx,
\qquad t>0.
\end{eqnarray}
They studied the relation between DCRE and well-known reliability measures. Other interesting
properties are given in a recent paper by Navarro et al.\ (2010). Baratpour (2010) studied the CRE of
first order statistics. A dynamic measure of discrimination between two lifetime distributions based on
CRE is introduced in Chamany and Baratpour (2014).  In order to pinpoint the age effect on
the information concerning the residual lifetime of a system, an analogous dynamic version of CRI,
called {\em dynamic cumulative residual inaccuracy} (DCRI) is defined as
\begin{eqnarray}\label{eq2.2}
\mathcal{C}H_{X,Y}(t)=-\int_{t}^{\infty}\frac{\overline F(x)}{\overline F(t)}\ln\frac{\overline G(x)}{\overline G(t)}dx
=-\int_{t}^{\infty}\overline F_t(x)\ln\overline G_t(x) dx,
\qquad t>0,
\end{eqnarray}
where $\overline F_t(x)=\frac{\overline F(x)}{\overline F(t)}$ and $\overline G_t(x)=\frac{\overline G(x)}{\overline G(t)}$,
$x>t$. When the two distributions coincide, the measure (\ref{eq2.2}) reduces to (\ref{eq2.1}).
Moreover, from Eqs.\ (\ref{eq5}) and (\ref{eq2.2}), $\lim_{t\to 0^+} \mathcal{C}H_{X,Y}(t)=\mathcal{C}H_{X,Y}$.
\par\hspace*{.2in}
Let us study some properties and bounds of CRI  in terms of CRE and  means of $X$ and $Y$.
\begin{p1}\label{p2.1}
If $X$ and $Y$ are two nonnegative random variables with finite means $E(X)$ and $E(Y)$, respectively, then \\
$(i)~\mathcal{C}H_{X,Y}\geqslant \varepsilon(X)+E(X)\ln\frac{E(X)}{E(Y)}$, \\
$(ii)~\mathcal{C}H_{X,Y}\geqslant \varepsilon(X)+\left[E(X)-E(Y)\right]$.
\end{p1}
Proof: The proof is immediate on using the log-sum inequality and the inequality
$a\ln\frac{a}{b}\geqslant a-b,~\forall a,b>0$.
$\hfill\square$
\par
\hspace*{.2in}
We recall that a random variable $X$ is said to be less than $Y$ in the usual stochastic order, written
as $X\leqslant_{st}Y$, if $\overline F(x)\leqslant\overline G(x)$ (see Shaked and Shanthikumar, 2007).
\begin{p1}\label{p2.2}
Let $X$ and $Y$ be two nonnegative random variables. \\
{\em (i)} If $X\leqslant_{st}~Y$, then $\mathcal{C}H_{X,Y}\leqslant {\rm min}\{\varepsilon(X),~\varepsilon(Y)\}$.\\
{\em (ii)} If $X\geqslant_{st}Y$, then $\mathcal{C}H_{X,Y}\geqslant {\rm max}\{\varepsilon(X),~\varepsilon(Y)\}$.
$\hfill\square$
\end{p1}
\hspace*{.2in}
The following proposition will be used to prove the upcoming theorem. The proof is easy and hence omitted.
\begin{p1}\label{p2.3}
Let $X,~Y$ and $Z$ be nonnegative random variables. \\
{\em (i)}~If $Y\leqslant_{st}Z$ then $\mathcal{C}H_{X,Y}\geqslant\mathcal{C}H_{X,Z}$. \\
{\em (ii)}~If~$X\leqslant_{st}Y$ then $\mathcal{C}H_{X,Z}\leqslant\mathcal{C}H_{Y,Z}$. $\hfill\square$
\end{p1}
On using the above result we have the following theorem.
\begin{t1}\label{t2.1}
Let $X,~Y$ and $Z$ be nonnegative random variables. If $X\leqslant_{st}Z\leqslant_{st}Y$, then
$$
 \mathcal{C}H_{Y,X}\geqslant {\rm max}\{\mathcal{C}H_{Y,Z},~\mathcal{C}H_{Z,X}\}.
 \qquad \qquad \square
$$
\end{t1}
\hspace*{.2in}
The following corollary involves mixture distributions, which play an important role in many branches
of statistics and applied probability. The proof follows from Theorem \ref{t2.1}, and from the fact that
if $X\leqslant_{st}Y$ and $Z$ is a mixture of $X$ and $Y$, then $X\leqslant_{st}Z\leqslant_{st}Y$.
\begin{c1}
Let $X$ and $Y$ be nonnegative random variables, and let $Z$ be a mixture of $X$ and $Y$.
If $X\leqslant_{st}Y$, then
$\mathcal{C}H_{Y,X}\geqslant {\rm max}\{\mathcal{C}H_{Y,Z},~\mathcal{C}H_{Z,X}\}$.
$\hfill\square$
\end{c1}
\hspace*{.2in}
We now show that the triangle inequality for the CRI is satisfied under some conditions.
\begin{t1}\label{t2.2} Let $X,~Y$ and $Z$ be nonnegative random variables with survival functions
$\overline F,~ \overline G$ and $\overline H$, respectively. If $(i)$ $X\leqslant_{st}Y$
and $Z\leqslant_{st}Y$ or $(ii)$ $Y\leqslant_{st}X$ and $Y\leqslant_{st}Z$,
then $$\mathcal{C}H_{X,Y}+\mathcal{C}H_{Y,Z}\geqslant\mathcal{C}H_{X,Z}.$$
\end{t1}
Proof: Let us assume that $(i)$ or $(ii)$ holds. Then
$\mathcal{C}H_{X,Y}+\mathcal{C}H_{Y,Z}\geqslant\varepsilon(Y)+\mathcal{C}H_{X,Z}$.
Hence, the result follows by noting that $\varepsilon(Y)$ is nonnegative.
$\hfill\square$\\
\hspace*{.2in}
Now we obtain similar results for the DCRI. Note that (\ref{eq2.2}) can be rewritten as
$$
 \mathcal{C}H_{X,Y}(t)=\delta_F(t)\ln\overline G(t)-\frac{1}{\overline F(t)}\int_t^\infty\overline F(x)\ln\overline G(x)dx,
 \qquad t>0,
$$
where $\delta_F(t)=E[X-t|X>t]=\frac{1}{\overline F(t)}\int_t^{\infty}\overline F(x)dx$, $t>0$,
is the mean residual life of $X$, and $\delta_G(t)$ is similarly defined for $Y$.
\begin{r1}
CRI and DCRI need not exist for all distributions. For example, let $X$ follow Pareto-I distribution with
$\overline F(x)=x^{-1},~x\geqslant1,$ and let $Y$ be standard exponential. It is easy to prove
that  $\mathcal{C}H_{X,Y}$ and $\mathcal{C}H_{X,Y}(t)$ are not finite. Thus, all the results discussed
here are based on the assumption that CRI and DCRI are finite.
$\hfill\square$
\end{r1}
\hspace*{.2in}
Differentiating (\ref{eq2.2}) with respect to $t$, we get
$$
 \frac{d}{dt}\mathcal{C}H_{X,Y}(t)=\lambda_F(t)\mathcal{C}H_{X,Y}(t)-\lambda_G(t)\delta_G(t),
$$
where $\lambda_F$ and $\lambda_G$ are hazard rates of $X$ and $Y$, respectively.
Therefore, DCRI is increasing (decreasing) in $t$ iff
$$
 \mathcal{C}H_{X,Y}(t)\geqslant(\leqslant)\frac{\lambda_G(t)}{\lambda_F(t)}\delta_G(t).
$$
In analogy with DCRE (ref.\ Examples 3.6 and 3.7 of Navarro et al., 2010), DCRI may be increasing and decreasing
in $t$. To see that not all distributions are monotone in terms of DCRI consider the following example.
\begin{ex}\label{ex2.1}
Let $X$ have survival function
\begin{equation*}
\overline F(x)=\left\{\begin{array}{ll}
1, &  x\leqslant3\\
e^{6-2x}, & 3<x<4\\
e^{2-x}, & x\geqslant4
\end{array}\right.
\end{equation*}
and for $Y$, $\overline G(x)=\sqrt{\overline F(x)}$. Then the dynamic cumulative residual inaccuracy is
\begin{equation*}
\mathcal{C}H_{X,Y}(t)
=\left\{\begin{array}{ll}
\frac{e^{2t-6}}{4}\left[(2t-9)e^{-2}-(2t-7)\right]-\left(\frac{t-5}{2}\right)e^{t-4}, & t\leqslant3\\
\frac{1}{4}\left[(2t-9)e^{2t-8}+1\right]-\left(\frac{t-5}{2}\right)e^{t-4}, & 3<t<4\\
\frac{1}{2}, & t\geqslant4
\end{array}\right.
\end{equation*}
Figure \ref{fig2.1} shows that $\mathcal{C}H_{X,Y}(t)$ is not monotone. $\hfill\square$
\begin{figure}[t]
\centering
\includegraphics[width=6.5 cm,keepaspectratio]{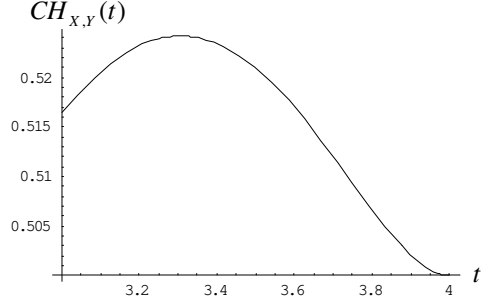}
\vspace{-.1cm}
\caption{Plot of $\mathcal{C}H_{X,Y}(t)$ for $t\in(3,4)$ (Example \ref{ex2.1}).}\label{fig2.1}
\end{figure}
\end{ex}
\hspace*{.2in}
Let us now discuss the effect of linear transformations on DCRI.
\begin{t1}\label{theor:ltdcri}
Let $X$ and $Y$ be nonnegative random variables $X$ and $Y$. For all $a>0$ and $0<b<t$ we have
$$
 \mathcal{C}H_{aX+b,aY+b}(t)=a\mathcal{C}H_{X,Y}\left(\frac{t-b}{a}\right).\qquad \hfill\square
$$

\end{t1}
\hspace*{.2in}
Classification of distributions with respect to ageing properties is a popular theme in reliability theory.
We recall the following classes of distributions which arise in the study of replacement and maintenance
policies: A nonnegative random variable $X$ is said to be
\\
$(i)$ new better than used (NBU) [new worse than used (NWU)] if
$\overline F(x+t)\leqslant[\geqslant]\overline F(x)\overline F(t)$, for all $x,t>0$;
\\
$(ii)$ new better than used in expectation (NBUE) [new worse than used  in expectation (NWUE)]
if $\delta_F(t)\leqslant[\geqslant]\delta_F(0)=E(X)$, for all $t>0$.
\\
See Barlow and Proschan (1975) for the details of some other concepts of ageing properties.
\par
\hspace*{.2in}
In the following we obtain lower bounds for DCRI. The proof follows on the same line of
Proposition \ref{p2.1} and hence is omitted.
\begin{p1}\label{p2.4}
Let $X$ and $Y$ be nonnegative random variables with finite means. Then, for $t>0$ \\
$(i)~\mathcal{C}H_{X,Y}(t)\geqslant\varepsilon(X;t)+\delta_F(t)\displaystyle\ln\left(\frac{\delta_F(t)}{\delta_G(t)}\right)$; \\
$(ii)~\mathcal{C}H_{X,Y}(t)\geqslant\varepsilon(X;t)+\left(E(X)-E(Y)\right)$ if $X$ and $Y$ are NWUE and NBUE, respectively.
\end{p1}
\hspace*{.2in}
We now find an upper bound for the difference between $\mathcal{C}H_{X,Y}$
and $\mathcal{C}H_{X,Y}(t)$.
\begin{p1}
For two nonnegative random variables $X$ and $Y$, if $X$ is NWU and $Y$ is NBU then
$$
 \mathcal{C}H_{X,Y}-\mathcal{C}H_{X,Y}(t)\leqslant\varepsilon(X)-\varepsilon(X;t), \qquad t>0.
$$
\end{p1}
Proof: On using the definitions of NWU and NBU, we have
$$
 \int_0^\infty\frac{\overline F(x+t)}{\overline F(t)}\ln\frac{\overline F(x+t)/\overline F(t)}{\overline G(x+t)/\overline G(t)}dx
 \geqslant\int_0^\infty\overline F(x)\ln\frac{\overline F(x)}{\overline G(x)}dx.
$$
Hence the result follows.
$\hfill\square$
\\
\hspace*{.2in}
In the following theorem, by using the concept of the hazard rate order, we obtain
bound of DCRI in terms of DCRE. Recall that a random variable $X$ is said to be smaller than $Y$
in hazard rate order, written as $X\leqslant_{hr}Y$, if $\lambda_F(t)\geqslant\lambda_G(t),~t\geqslant 0.$
\begin{p1}
Let $X$ and $Y$ be nonnegative random variables. \\
{\em (i)} If $X\leqslant_{hr}~Y$, then
$\mathcal{C}H_{X,Y}(t)\leqslant {\rm min}\{\varepsilon(X;t),~\varepsilon(Y;t)\}$, $t \geqslant 0$. \\
{\em (ii)} If $X\geqslant_{hr}Y$, then $\mathcal{C}H_{X,Y}(t)\geqslant {\rm max}\{\varepsilon(X;t),~\varepsilon(Y;t)\}$,
$t \geqslant 0$.
\end{p1}
Proof: The proof follows from (\ref{eq2.2}) and using the fact that $X\leqslant_{hr}Y$
is equivalent to $\overline F_t(x)\leqslant\overline G_t(x)$, for $x,t \geqslant 0$.
$\hfill\square$
\\
\hspace*{.2in}
The following result is on the same line of Proposition \ref{p2.3}.
\begin{p1}
Let $X,~Y$ and $Z$ be nonnegative random variables. \\
{\em (i)} If $Y\leqslant_{hr}Z$ then
$\mathcal{C}H_{X,Y}(t)\geqslant\mathcal{C}H_{X,Z}(t)$,  $t \geqslant 0$,\\
{\em (ii)} If $X\leqslant_{hr}Y$ then $\mathcal{C}H_{X,Z}(t)\leqslant\mathcal{C}H_{Y,Z}(t)$, $t \geqslant 0$.
$\hfill\square$
\end{p1}
\hspace*{.2in}
On using the above we have the following theorem.
\begin{t1}
Let $X,~Y$ and $Z$ be  nonnegative random variables. If $X\leqslant_{hr}Z\leqslant_{hr}~Y$,
then $\mathcal{C}H_{Y,X}(t)\geqslant {\rm max}\{\mathcal{C}H_{Y,Z}(t),~\mathcal{C}H_{Z,X}(t)\},$
$t\geqslant 0$.
$\hfill\square$
\end{t1}
\begin{c1}
Let $X$ and $Y$ be nonnegative random variables, and let $Z$ be a mixture of $X$ and $Y$.
If $X\leqslant_{hr}Y$, then
$\mathcal{C}H_{Y,X}(t)\geqslant {\rm max}\{\mathcal{C}H_{Y,Z}(t),~\mathcal{C}H_{Z,X}(t)\}$,
$t \geqslant 0$.
$\hfill\square$
\end{c1}
\hspace*{.2in}
The proportional hazards model (also known as Cox model) is largely employed in survival
analysis and statistics (see, for instance, Cox and Oakes, 1984). It refers to a pair of
nonnegative random variables $X$ and $Y$, whose survival functions are related by this relation:
\begin{equation}
 \overline F(x)=[\overline G(x)]^{\alpha}, \qquad x\geqslant 0, \qquad (\alpha>0,\;\; \alpha\neq 1).
 \label{eq:phmod}
\end{equation}
The following result is an immediate consequence of Eqs.\ (\ref{eq2.1}), (\ref{eq2.2}) and (\ref{eq:phmod}).
\begin{p1} \label{prop:rhm}
Let $X$ and $Y$ be nonnegative  random variables with reliability functions $\overline F(x)$ and
$\overline G(x)$, respectively, satisfying the proportional hazards model (\ref{eq:phmod}). Then,
$$
 \mathcal{C}H_{X,Y}(t)=\alpha\cdot \varepsilon(X;t), \qquad t \geq 0.
$$
\end{p1}
\hspace*{.2in}
We conclude this section by showing that the triangle inequality for $\mathcal{C}H_{X,Y}(t)$
is satisfied under stronger conditions than those of Theorem \ref{t2.2}. The proof is similar and
then omitted.
\begin{t1}\label{th:trinhr}
Let $X,~Y$ and $Z$ be nonnegative random variables with survival functions
$\overline F,~ \overline G$ and $\overline H$, respectively. If $(i)$ $X\leqslant_{hr}Y$
and $Z\leqslant_{hr}Y$, or $(ii)$ $Y\leqslant_{hr}X$ and $Y\leqslant_{hr}Z$, then
$$
 \mathcal{C}H_{X,Y}(t)+\mathcal{C}H_{Y,Z}(t)\geqslant\mathcal{C}H_{X,Z}(t),
 \qquad t \geqslant 0.
$$
\end{t1}
\section{Results on (dynamic) CPI}
Measure of uncertainty in past lifetime distribution plays an important role in the context of information theory,
forensic sciences, and other related fields. Suppose that a system or a component fails at time $t(>0)$. Then
Di Crescenzo and Longobardi (2009) proposed {\em dynamic cumulative past entropy} (DCPE)
based on CPE for the past lifetime distribution corresponding to the random variable $[X|X\leqslant t]$ as
\begin{eqnarray}\label{eq3.1}
\overline \varepsilon(X;t)=-\int_0^t\frac{F(x)}{F(t)}\ln\frac{F(x)}{F(t)}dx,
\qquad t>0.
\end{eqnarray}
They studied the monotonicity properties of this measure and certain bounds. Some other results on
DCPE are available in Navarro et al.\ (2010). It should be noted that the random variable $X_{(t)}=[X|X\leqslant t]$
has a nice application in economics, since it represents the income distribution of the poor for a poverty line $t$.
In analogy with (\ref{eq2.2}), we define the {\em dynamic cumulative past inaccuracy} (DCPI) as
\begin{eqnarray}\label{eq3.2}
\mathcal{C}\overline H_{X,Y}(t)=-\int_0^t\frac{F(x)}{F(t)}\ln\frac{G(x)}{G(t)}dx
=-\int_0^t F_t(x)\ln G_t(x) dx,\qquad t>0,
\end{eqnarray}
where $F_t(x)=\frac{F(x)}{F(t)}$ and $G_t(x)=\frac{G(x)}{G(t)}$, $0\leqslant x\leqslant t$. Now we study some
properties and bounds of CPI in analogy with CRI. The proofs are omitted. For some recent results on CPI
and empirical CPI based on suitable stochastic orderings, see Di Crescenzo and Longobardi (2013).
\begin{p1}
Let random variables $X$ and $Y$ take values in $[0,b]$ with $b$ finite. Then \\
{\em (i)}
$\mathcal{C}\overline H_{X,Y}\geqslant\overline \varepsilon(X)+\left(b-E(X)\right)\ln\left(\frac{b-E(X)}{b-E(Y)}\right)$; \\
{\em (ii)}
$\mathcal{C}\overline H_{X,Y}\geqslant\overline \varepsilon(X)+\left(E(Y)-E(X)\right)$;  \\
{\em (iii)}
if $X\leqslant_{st}Y$, then
$\mathcal{C}\overline H_{X,Y}\geqslant {\rm max}\{\overline \varepsilon(X),~\overline \varepsilon(Y)\}$; \\
{\em (iv)}
if $X\geqslant_{st}Y$, then
$\mathcal{C}\overline H_{X,Y}\leqslant {\rm min}\{\overline \varepsilon(X),~\overline \varepsilon(Y)\}$.
$\hfill\square$
\end{p1}
\begin{p1}
Let $X,~Y$ and $Z$ be random variables with finite support $[0,b]$. \\
{\em (i)} If $Y\geqslant_{st}Z$ then $\mathcal{C}\overline H_{X,Y}\geqslant\mathcal{C}\overline H_{X,Z}$. \\
{\em (ii)} If $X\geqslant_{st}Y$ then $\mathcal{C}\overline H_{X,Z}\leqslant\mathcal{C}\overline H_{Y,Z}$. \\
{\em (iii)} If $X\geqslant_{st}Z\geqslant_{st}Y$ then
$\mathcal{C}\overline H_{Y,X}\geqslant {\rm max}\{\mathcal{C}\overline H_{Y,Z},~\mathcal{C}\overline H_{Z,X}\}$.
$\hfill\square$
\end{p1}
\begin{c1} Let $X$ and $Y$ be random variables with finite support $[0,b]$, and let $Z$ be a mixture of $X$ and $Y$.
If $X\geqslant_{st}Y$, then
$\mathcal{C}\overline H_{Y,X}\geqslant {\rm max}\{\mathcal{C}\overline H_{Y,Z},~\mathcal{C}\overline H_{Z,X}\}$.
$\hfill\square$
\end{c1}
\hspace*{.2in}
The following theorem investigates the triangle inequality for $\mathcal{C}\overline H_{X,Y}$.
\begin{t1}
Let $X,~Y$ and $Z$ be nonnegative random variables with finite support $[0,b]$.
If $(i)$ $X\leqslant_{st}Y$ and $Z\leqslant_{st}Y$ or $(ii)$ $Y\leqslant_{st}X$ and $Y\leqslant_{st}Z$, then
$$
 \mathcal{C}\overline H_{X,Y}+\mathcal{C}\overline H_{Y,Z}\geqslant\mathcal{C}\overline H_{X,Z}.
 \qquad \qquad \square
$$
\end{t1}
\hspace*{.2in}
Now we consider analogous results for DCPI. Note that (\ref{eq3.2}) can be written as
$$
 \mathcal{C}\overline H_{X,Y}(t)=\ln G(t)m_F(t)-\frac{1}{F(t)}\int_0^tF(x)\ln G(x)dx,
 \qquad t>0,
$$
where $m_F(t)=E[t-X|X\leqslant t]$ is the expected inactivity time of $X$, and $m_G(t)$ is similarly
defined for $Y$. An alternative expression to (\ref{eq3.2}) is provided hereafter. We recall that an
analogous expression for (\ref{eq3.1}) is given in Remark 5.1 of Di Crescenzo and Longobardi (2009).
\begin{p1}\label{p3.4}
For two absolutely continuous nonnegative random variables $X$ and $Y$,
$$
 {\cal C}\overline H_{Y,X}(t)=E[\tau_F^{(2)}(Y,t)|Y\leqslant t], \qquad t>0,
$$
where
$$
  \tau_F^{(2)}(x,t)=-\int_x^t \ln \frac{F(u)}{F(t)}du, \qquad 0\leqslant x<t.
$$
\end{p1}
Proof: Using Fubini's theorem, for $t>0$, we have
\begin{eqnarray*}
  E[\tau_F^{(2)}(Y,t)|Y\leqslant t]
  &=& -\int_0^t \frac{g(u)}{G(t)}\left(\int_u^t
\ln\frac{F(x)}{F(t)}dx\right)du \\
  &=& -\int_0^t \frac{1}{G(t)}\left(\int_0^x g(u)du\right)\ln\frac{F(x)}{F(t)}dx
  ={\cal C}\overline H_{Y,X}(t).~~\qquad \qquad \square
\end{eqnarray*}
\begin{r1}
Differentiating (\ref{eq3.2}) with respect to $t$, we get
$$
 \frac{d}{dt}\mathcal{C}\overline H_{X,Y}(t)=\phi_G(t)m_F(t)-\phi_F(t)\mathcal{C}\overline H_{X,Y}(t),
$$
where $\phi_F$ and $\phi_G$ are reversed hazard rates of $X$ and $Y$, respectively. Therefore,
DCPI is increasing (decreasing) in $t$ iff
$$
 \mathcal{C}\overline H_{X,Y}(t)\leqslant(\geqslant)\frac{\phi_G(t)}{\phi_F(t)}m_F(t).
$$
\end{r1}
\hspace*{.2in}
The following example shows that DCPI is not monotone for all distributions.
\begin{ex}\label{ex3.1}
Let $X$ and $Y$ have distribution functions
\begin{equation*}
F(x)=\left\{\begin{array}{ll}
\exp\{-1/2-1/x\}, & 0<x\leqslant1 \\
\exp\{-2+x^2/2\}, &1<x\leqslant2\\
1, &x\geqslant2
\end{array}
\right.
\quad {\rm and} \quad G(x)
=\left\{\begin{array}{ll}
x^2/4, &0<x\leqslant 2 \\
1, & x\geqslant 2.
\end{array}
\right.
\end{equation*}
Then, for $t\geqslant2$,
$$
 \mathcal{C}\overline H_{X,Y}(t)=-2\left[\int_0^1e^{1/t-1/x}\ln(x/t)dx+\int_1^2e^{(x^2-t^2)/2}\ln(x/t)dx\right],
$$
which is not monotone as shown in Figure \ref{fig3.1}.
$\hfill\square$
\begin{figure}[t]
\centering
\includegraphics[width=6.5 cm,keepaspectratio]{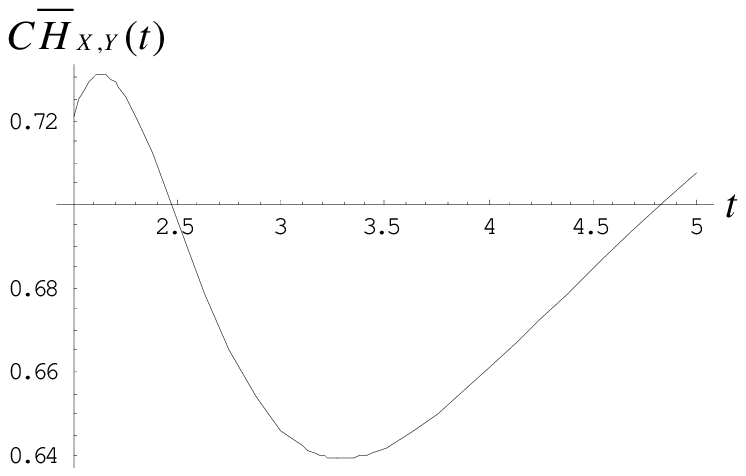}
\vspace{-.1cm}
\caption{Plot of $\mathcal{C}\overline H_{X,Y}(t)$ for $t\in(2,5)$ (Example \ref{ex3.1}).}\label{fig3.1}
\end{figure}
\end{ex}
\hspace*{.2in}
In analogy with Theorem \ref{theor:ltdcri} we now discuss the effect of linear transformations on DCPI.
\begin{t1}
For two nonnegative random variables $X$ and $Y$, for all $a>0$ and $0<b<t$,
$$
 \mathcal{C}\overline H_{aX+b,aY+b}(t)=a\mathcal{C}\overline H_{X,Y}\left(\frac{t-b}{a}\right).
 \qquad \qquad  \square
$$
\end{t1}
\hspace*{.2in}
Now we show an identity for the DCPI and DCRI of symmetric distributions.
The proof follows from (\ref{eq3.2}) and (\ref{eq2.2}).
\begin{t1}
Let $X$ and $Y$ be random variables with finite support $[0,b]$, and
symmetric with respect to $b/2$, i.e., $F(x)=\overline F(b-x)$ and $G(x)=\overline G(b-x)$
for $0\leqslant x\leqslant b$. Then,
$$
 \mathcal{C}\overline H_{X,Y}(t)=\mathcal{C}H_{X,Y}(b-t).~~\qquad \qquad \square
$$
\end{t1}
\hspace*{.2in}
The following properties and bounds of DCPI are analogous to the same results for CPI
and thus the proof is omitted. Recall that a random variable $X$ is said to be smaller than $Y$
in reversed hazard rate order, written as $X\leqslant_{rh}Y$, if $\phi_F(t)\leqslant\phi_G(t),~t\geqslant0$,
or equivalently, $X_{(t)}\leqslant_{st}Y_{(t)}$ for all $t\geqslant0$.
\begin{p1}
For two nonnegative random variables $X$ and $Y$, for $t\geqslant0$,
\begin{itemize}
\item[$\bullet$] $\mathcal{C}\overline H_{X,Y}(t)\geqslant\overline\varepsilon(X;t)+m_F(t)\ln\left(\frac{m_F(t)}{m_G(t)}\right)$;
\item[$\bullet$] $\mathcal{C}\overline H_{X,Y}(t)\geqslant\overline\varepsilon(X;t)+\left(m_F(t)-m_G(t)\right)$;
\item[$\bullet$] $\mathcal{C}\overline H_{X,Y}(t)\leqslant{\rm min}\{\overline\varepsilon(X;t),~\overline\varepsilon(Y;t)\}$, if $X\geqslant_{rh}Y$;
\item[$\bullet$] $\mathcal{C}\overline H_{X,Y}(t)\geqslant{\rm max}\{\overline\varepsilon(X;t),~\overline\varepsilon(Y;t)\}$, if $X\leqslant_{rh}Y$.
\end{itemize}
\end{p1}
\begin{p1}
Let $X,~Y$ and $Z$ be nonnegative random variables. Then, for $t\geqslant0$,
\begin{itemize}
\item[$\bullet$] $\mathcal{C}\overline H_{X,Y}(t)\geqslant\mathcal{C}\overline H_{X,Z}(t)$, if $Y\geqslant_{rh}Z$;
\item[$\bullet$] $\mathcal{C}\overline H_{X,Z}(t)\leqslant\mathcal{C}\overline H_{Y,Z}(t)$, if $X\geqslant_{rh}Y$;
\item[$\bullet$] $\mathcal{C}\overline H_{Y,X}(t)\geqslant {\rm max}\{\mathcal{C}\overline H_{Y,Z}(t),
~\mathcal{C}\overline H_{Z,X}(t)\}$, if $X\geqslant_{rh}Z\geqslant_{rh}Y$.
\end{itemize}
\end{p1}
\begin{p1}
Let $X$ and $Y$ be nonnegative random variables and let $Z$ be a mixture of $X$ and $Y$.
If $X\geqslant_{rh}Y$, then
$$
 \mathcal{C}\overline H_{Y,X}(t)\geqslant {\rm max}\{\mathcal{C}\overline H_{Y,Z}(t),
 ~\mathcal{C}\overline H_{Z,X}(t)\}.~~\qquad \qquad \hfill\square
$$
\end{p1}
\begin{t1}
Let $X$ and $Y$ be absolutely continuous nonnegative random variables satisfying $X\leqslant_{rh}Y$ and $\mu_X(t)<\mu_Y(t)$ for all $t>0$, where $\mu_X(t)=E[X_{(t)}]$, and $\mu_Y(t)$ is similarly defined for $Y_{(t)}$.
If both ${\cal C}\overline H_{Y,X}(t)$ and $\bar \varepsilon (X;t)$ are finite, then for all $t>0$
$$
{\cal C}\overline H_{Y,X}(t)
  =\bar \varepsilon (X;t)
  +E[\dot \tau_F^{(2)}(Z_t,t)] \{\mu_Y(t)-\mu_X(t)\},
$$
where $\dot \tau_F^{(2)}(z,t)=(d/dz)\tau_F^{(2)}(z,t)$ and $Z_t=\Psi(X_{(t)},Y_{(t)})$ is an absolutely continuous nonnegative random variable with probability density  (cf.\ Proposition 3.1 of Di Crescenzo, 1999)
$$
 f_{Z_t}(x)=\frac{1}{\mu_Y(t)-\mu_X(t)}\left[\frac{F(x)}{F(t)}-\frac{G(x)}{G(t)}\right],\qquad 0<x<t.
$$
\end{t1}
Proof:
On using Theorem 4.1 of Di Crescenzo (1999), the proof is an immediate consequence of Proposition \ref{p3.4},
and Remark 5.1 of Di Crescenzo and Longobardi (2009).~$\hfill\square$\\
\hspace*{.2in}
Dual to the model considered in Eq.\ (\ref{eq:phmod}), the proportional reversed hazards model
refers to the distribution functions of nonnegative random variables $X$ and $Y$ that are
related by the following relation (see for instance Di Crescenzo, 2000, Gupta and Gupta, 2007,
Sankaran and Gleeja, 2008):
\begin{equation}
 F(x)=[G(x)]^{\theta}, \qquad x\geqslant 0, \qquad (\theta >0,\;\; \theta\neq 1).
 \label{eq:prhmod}
\end{equation}
Similarly to Proposition \ref{prop:rhm}, the following result follows from Eqs.\
(\ref{eq3.1}), (\ref{eq3.2}) and (\ref{eq:prhmod}).
\begin{p1}
Let $X$ and $Y$ be nonnegative  random variables satisfying the proportional reversed hazards model. Then,
$$
 \mathcal{C}\overline H_{X,Y}(t)=\theta\cdot \overline \varepsilon(X;t), \qquad t \geq 0.
$$
\end{p1}
\hspace*{.2in}
We conclude this section by showing that the triangle inequality is also satisfied for DCPI under suitable
conditions, similarly to Theorem \ref{th:trinhr}.
\begin{t1}
Let $X,~Y$ and $Z$ be three nonnegative random variables.
If $(i)$ $X\leqslant_{rh}Y$ and $Z\leqslant_{rh}Y$ or,
$(ii)$ $Y\leqslant_{rh}X$ and $Y\leqslant_{rh}Z$, then
$$
 \mathcal{C}\overline H_{X,Y}(t)+\mathcal{C}\overline H_{Y,Z}(t)\geqslant\mathcal{C}\overline H_{X,Z}(t),
 \qquad t>0.
$$
\end{t1}
\section{Some properties of interval CRI and CPI}
Most of the real life observations are truncated in nature. In information theory and reliability, one has
information about the lifetime of an individual between two time instants. Thus, an individual whose
event time is not in this interval is not observed. For example, in insurance, claim time of a policy
holder is doubly truncated between starting date and maturity date of the policy. Doubly truncated
data play an important role in the statistical analysis of astronomical observations also.
These reasons motivate us to consider the inaccuracy measure of two nonnegative absolutely
continuous doubly truncated random variables
$[X|t_1\leqslant X\leqslant t_2]$ and $[Y|t_1\leqslant Y\leqslant t_2]$ where
$(t_1,t_2)\in D:=\{(u,v)\in\mathbb{R}_+^2:F(u)<F(v) ~{\rm and}~G(u)<G(v)\}$. Then, the
{\em interval inaccuracy measure} of $X$ and $Y$ in the interval $(t_1,t_2)$ is given by
\begin{eqnarray}\label{eq4.1}
H_{X,Y}(t_1,t_2)=-\int_{t_1}^{t_2}
\frac{f(x)}{F(t_2)-F(t_1)}\ln\frac{g(x)}{G(t_2)-G(t_1)}dx.
\end{eqnarray}
Various aspects of (\ref{eq4.1}) have been discussed in Kundu and Nanda (2014). When $g(x)=f(x)$,
we obtain interval entropy of $X$ in $(t_1,t_2)$ studied by Sunoj et al. (2009) and Misagh and Yari
(2011, 2012), among others. Recently,  for doubly truncated random variables Khorashadizadeh et al.\ (2013)
introduced the concepts of {\em interval cumulative residual entropy} (ICRE) as
\begin{eqnarray}\label{eq4.2}
\varepsilon(X;t_1,t_2)=-\int_{t_1}^{t_2}
\frac{\overline F(x)}{\overline F(t_1)-\overline F(t_2)}\ln\frac{\overline F(x)}{\overline F(t_1)-\overline F(t_2)}dx,
\end{eqnarray}
and {\em interval cumulative past entropy} (ICPE) as
\begin{eqnarray}\label{eq4.3}
\overline\varepsilon(X;t_1,t_2)=-\int_{t_1}^{t_2}
\frac{F(x)}{F(t_2)-F(t_1)}\ln\frac{F(x)}{F(t_2)-F(t_1)}dx.
\end{eqnarray}
They studied several properties of (\ref{eq4.2}) and (\ref{eq4.3}), extending the results for DCRE and DCPE.
Similarly, for $(t_1,t_2)\in D$ we define the {\em interval cumulative residual inaccuracy} (ICRI):
\begin{eqnarray}\label{eq4.4}
\mathcal{IC}H_{X,Y}(t_1,t_2)=-\int_{t_1}^{t_2}
\frac{\overline F(x)}{\overline F(t_1)-\overline F(t_2)}\ln\frac{\overline G(x)}{\overline G(t_1)-\overline G(t_2)}dx
\end{eqnarray}
and the {\em interval cumulative past inaccuracy} (ICPI):
\begin{eqnarray}\label{eq4.5}
\mathcal{IC}\overline H_{X,Y}(t_1,t_2)=-\int_{t_1}^{t_2}
\frac{F(x)}{F(t_2)-F(t_1)}\ln\frac{G(x)}{G(t_2)-G(t_1)}dx.
\end{eqnarray}
Clearly, $\mathcal{IC}H_{X,Y}(t_1,\infty)$ is the DCRI and $\mathcal{IC}\overline H_{X,Y}(0,t_2)$
is the DCPI as defined in (\ref{eq2.2}) and (\ref{eq3.2}), respectively.
We remark that the ICRI can alternatively be written as
\begin{eqnarray*}\mathcal{IC}H_{X,Y}(t_1,t_2)&=&-\frac{1}{\overline F(t_1)-\overline F(t_2)}\int_{t_1}^{t_2}\overline F(x)\ln\overline G(x)dx\\
&&+\ln\{\overline G(t_1)-\overline G(t_2)\}\left[m_X(t_1,t_2)+\frac{t_2\overline F(t_2)-t_1\overline F(t_1)}{\overline F(t_1)-\overline F(t_2)}\right],
\end{eqnarray*}
where $m_X(t_1,t_2)=E[X|t_1\leqslant X\leqslant t_2]$ is the general conditional mean (GCM) of $X$.
Note that the above integral in the right-hand-side has the following nice probabilistic meaning:
\begin{eqnarray*}
-\frac{1}{\overline F(t_1)-\overline F(t_2)}\int_{t_1}^{t_2}
\overline F(x)\ln \overline G(x) dx
\!\! &=& \!\!  -\frac{1}{\overline F(t_1)-\overline
F(t_2)}\int_{t_1}^{t_2}f(u)\left[ \int_{t_1}^{u} \ln \overline G(x)
dx\right]du \\
&&  \!\! -\frac{1}{\overline F(t_1)-\overline
F(t_2)}\int_{t_2}^{\infty}f(u)\left[ \int_{t_1}^{t_2} \ln \overline
G(x) dx\right]du  \\
&=&  \!\! \frac{\overline F(t_2)}{\overline F(t_1)-\overline
F(t_2)}\Lambda_Y^{(2)}(t_1,t_2)
+E\left[\Lambda_Y^{(2)}(t_1,X)|t_1\leqslant X\leqslant t_2\right],
\end{eqnarray*}
where we have set, for $0\leq a<b$,
$$
 \Lambda_Y^{(2)}(a,b):=-\int_a^b \ln \overline G(x)dx=\int_a^b dx \int_0^{x} \lambda_G(u)du.
$$
Similarly, the ICPI can also alternatively be written as
\begin{eqnarray*}\mathcal{IC}\overline H_{X,Y}(t_1,t_2) \!\!&=& \!\!-\frac{1}{F(t_2)-F(t_1)}\int_{t_1}^{t_2}F(x)\ln G(x)dx\\
&& \!\!+\ln\{G(t_2)-G(t_1)\}\left[m_X(t_1,t_2)+\frac{t_2F(t_2)-t_1F(t_1)}{F(t_2)-F(t_1)}\right]\\
&=& \!\! \frac{F(t_1)}{F(t_2)-F(t_1)}T_Y^{(2)}(t_1,t_2)
+E\left[T_Y^{(2)}(X,t_2)|t_1\leqslant X\leqslant t_2\right]\\
&& \!\! +\ln\{G(t_2)-G(t_1)\}\left[m_X(t_1,t_2)+\frac{t_2F(t_2)-t_1F(t_1)}{F(t_2)-F(t_1)}\right],
\end{eqnarray*}
where
$$
 T_Y^{(2)}(a,b):=-\int_a^b \ln G(x)dx=\int_a^b dx \int_x^{\infty} \phi_G(u)du.
$$
\par\hspace*{.2in}
Now we study some properties of ICRI and ICPI including monotonicity and  bounds. Some of the results
presented here are similar, but more general, to corresponding results of Khorashadizadeh et al.\ (2013).
We first give definition of general failure rate (GFR). For more details on GCM and GFR we refer to
Navarro and Ruiz (1996) and Sunoj et al.\ (2009).
\begin{d1}
The GFR functions of a doubly truncated random variable $[X|t_1<X<t_2]$ are given by
$h^X_1(t_1,t_2)=\frac{f(t_1)}{F(t_2)-F(t_1)}$ and $h^X_2(t_1,t_2)=\frac{f(t_2)}{F(t_2)-F(t_1)}$. For the
random variable $[Y|t_1<Y<t_2]$ the GFRs $h^Y_1(t_1,t_2)$ and $h^Y_2(t_1,t_2)$ are defined similarly.
$\hfill\square$
\end{d1}
\hspace*{.2in}
On differentiating (\ref{eq4.4}) with respect to $t_1$, we get
\begin{eqnarray}\label{eq4.6}
\frac{\partial}{\partial t_1}\mathcal{IC}H_{X,Y}(t_1,t_2) \!\!\! &=&  \!\!\! h_1^X(t_1,t_2)
\left[\mathcal{IC}H_{X,Y}(t_1,t_2)-\frac{h_1^Y(t_1,t_2)}{h_1^X(t_1,t_2)}\left(m_X(t_1,t_2)+\frac{t_2\overline F(t_2)-t_1\overline F(t_1)}{\overline F(t_1)-\overline F(t_2)}\right)\nonumber\right.\\
&&\left.+\ln\left(\frac{\overline G(t_1}{\overline G(t_1)-\overline G(t_2)}\right)^{\frac{1}{\lambda_F(t_1)}}\right].
\end{eqnarray}
\hspace*{.2in}
The following theorem shows that there exist no nonnegative random variables for which ICRI
is increasing over the domain $D$. We omit the proof, being similar to that of Theorem 2.2 of
Khorashadizadeh et al.\ (2013).
\begin{t1}\label{t4.1}
If $X$ and $Y$ are nonnegative non-degenerate random variables then the ICRI cannot be
increasing with respect to $t_1$, for fixed $t_2$, where $(t_1,t_2)\in D$.
\end{t1}
%
%
\par\hspace*{.2in}
It should be noted that in special case $\mathcal{IC}H_{X,Y}(t_1,\infty)=\mathcal{C}H_{X,Y}(t_1)$
may be an increasing and a decreasing function of $t_1$.
\par\hspace*{.2in}
In the following theorem we obtain lower and upper bounds for ICRI.
\begin{t1}
Let $X$ and $Y$ be absolutely continuous nonnegative random variables, and let $(t_1,t_2)\in D$. Then, \\
{\em (i)}
$$
 \mathcal{IC}H_{X,Y}(t_1,t_2)\geqslant(t_1-t_2)\frac{h_1^X(t_1,t_2)}{\lambda_F(t_1)}\ln\left(\frac{h_1^Y(t_1,t_2)}   {\lambda_G(t_1)}\right);
$$
{\em (ii)}
if ICRI is decreasing in $t_1$, for fixed $t_2$,  then
$$
 \mathcal{IC}H_{X,Y}(t_1,t_2)\leqslant \frac{h_1^Y(t_1,t_2)}{h_1^X(t_1,t_2)}
 \left(m_X(t_1,t_2)+\frac{t_2\overline F(t_2)- t_1\overline F(t_1)}{\overline F(t_1)-\overline F(t_2)}\right)
 -\ln\left(\frac{h_1^Y(t_1,t_2)}{\lambda_G(t_1)}\right)^{\frac{1}{\lambda_F(t_1)}};
$$
{\em (iii)}
if $X$ and $Y$ have increasing (decreasing) hazard rates, then
$$
 \mathcal{IC}H_{X,Y}(t_1,t_2)\geqslant(\leqslant)\frac{1}{\lambda_F(t_1)}\left(H_{X,Y}(t_1,t_2)+\ln\lambda_G(t_1)\right),
$$
where $H_{X,Y}(t_1,t_2)$ is the interval inaccuracy measure defined in (\ref{eq4.1}).
$\hfill\square$
\end{t1}
\hspace*{.2in}
In the following theorem, the relationship between ICRI and ICRE is presented. The proof follows
on using the inequality $a\ln\frac{a}{b}\geqslant a-b,~\forall a,b>0$.
\begin{t1}\label{t4.5}
Let $X$ and $Y$ be two absolutely continuous nonnegative random variables and $(t_1,t_2)\in D$, then
$$
 \mathcal{IC}H_{X,Y}(t_1,t_2)\geqslant\varepsilon(X;t_1,t_2)+m_X(t_1,t_2)-m_Y(t_1,t_2)
 +\frac{t_2\overline F(t_2)-t_1\overline F(t_1)}{\overline F(t_1)-\overline F(t_2)}
 -\frac{t_2\overline G(t_2)-t_1\overline G(t_1)}{\overline G(t_1)-\overline G(t_2)}.
$$
\end{t1}
\hspace*{.2in}
The following properties and bounds for ICPI are similar to Theorem \ref{t4.1}-\ref{t4.5}.
\begin{r1}
For two absolutely continuous nonnegative random variables $X$ and $Y$ and $(t_1,t_2)\in D$, we have
\begin{itemize}
\item[$\bullet$] $\mathcal{IC}\overline H_{X,Y}(t_1,t_2)$ cannot be a decreasing function of $t_2$, for any fixed $t_1$;
\item[$\bullet$] $\mathcal{IC}\overline H_{X,Y}(t_1,t_2)\geqslant(t_1-t_2)\frac{h_2^X(t_1,t_2)}{\phi_F(t_2)}\ln\left(\frac{h_2^Y(t_1,t_2)}{\phi_G(t_2)}\right);$
\item[$\bullet$] $\mathcal{IC}\overline H_{X,Y}(t_1,t_2)$ is increasing in $t_2$, for fixed $t_1$, if and only if
$$\mathcal{IC}\overline H_{X,Y}(t_1,t_2)\leqslant \frac{h_2^Y(t_1,t_2)}{h_2^X(t_1,t_2)}\left(\frac{t_2F(t_2)-t_1F(t_1)}{F(t_2)-F(t_1)}-m_X(t_1,t_2)\right)-
\ln\left(\frac{h_2^Y(t_1,t_2)}{\phi_G(t_2)}\right)^{\frac{1}{\phi_F(t_2)}};$$
\item[$\bullet$] $\mathcal{IC}\overline H_{X,Y}(t_1,t_2)\geqslant\frac{1}{\phi_F(t_2)}\left(H_{X,Y}(t_1,t_2)+\ln\phi_G(t_2)\right),$ if $\phi_F,~\phi_G$ are decreasing functions;
\item[$\bullet$] $\mathcal{IC}\overline H_{X,Y}(t_1,t_2)\geqslant\overline\varepsilon(X;t_1,t_2)+m_Y(t_1,t_2)-m_X(t_1,t_2)+\frac{t_2F(t_2)-t_1F(t_1)}{F(t_2)-F(t_1)} -\frac{t_2G(t_2)-t_1G(t_1)}{G(t_2)-G(t_1)}.$
\end{itemize}
\end{r1}
\hspace*{.2in}
Now we discuss the effect of monotonic transformation on ICRI.
\begin{t1}\label{t4.6}
Let $X$ and $Y$ be absolutely continuous nonnegative random variables, and let $\varphi(\cdot)$ be
an increasing function on $[0,\infty)$. If $a\leqslant\varphi'\leqslant b,~a,b>0$, where $\varphi'$ is the
derivative of $\varphi$, then
$$
 b\cdot \mathcal{IC}H_{X,Y}(\varphi^{-1}(t_1),\varphi^{-1}(t_2))\leqslant
 \mathcal{IC}H_{\varphi(X),\varphi(Y)}(t_1,t_2)\leqslant a\cdot \mathcal{IC}H_{X,Y}(\varphi^{-1}(t_1),\varphi^{-1}(t_2)),
$$
and $\mathcal{IC}H_{bX,bY}(t_1,t_2)=b\cdot \mathcal{IC}H_{X,Y}(t_1/b,t_2/b)$.
If $\varphi$ is decreasing with $a\leqslant-\varphi'\leqslant b,~a,b>0$, then
$$
 b\cdot \mathcal{IC}\overline H_{X,Y}(\varphi^{-1}(t_2),\varphi^{-1}(t_1))
 \leqslant\mathcal{IC}H_{\varphi(X),\varphi(Y)}(t_1,t_2)\leqslant
 a\cdot \mathcal{IC}\overline H_{X,Y}(\varphi^{-1}(t_2),\varphi^{-1}(t_1)).
$$
\end{t1}
Proof: From (\ref{eq4.4}), if $\varphi$ is an increasing function we have
$$
 \mathcal{IC}H_{\varphi(X),\varphi(Y)}(t_1,t_2)
  =-\int_{\varphi^{-1}(t_1)}^{\varphi^{-1}(t_2)}\varphi'(y)\frac{\overline F(y)}{\overline F(\varphi^{-1}(t_1))-\overline F(\varphi^{-1}(t_2))}
  \ln\frac{\overline G(y)}{\overline G(\varphi^{-1}(t_1))-\overline G(\varphi^{-1}(t_2))}dy.
$$
Therefore the result follows on using $a\leqslant\varphi'\leqslant b,$ and later on taking $\varphi(x)=bx$, in particular.
When $\varphi$ is a decreasing function the proof proceeds similarly.
The rest of the proof follows from (\ref{eq4.5}) on using $a\leqslant-\varphi'\leqslant b.
$~$\hfill\square$
\begin{r1}\label{r4.2}
Let $X$ and $Y$ be absolutely continuous nonnegative random variables, and let $\varphi(\cdot)$ be
an increasing function on $[0,\infty)$. If $a\leqslant\varphi'\leqslant b,~a,b>0$, then
$$
 b\cdot \mathcal{IC}\overline H_{X,Y}(\varphi^{-1}(t_1),\varphi^{-1}(t_2))\leqslant\mathcal{IC}\overline H_{\varphi(X),\varphi(Y)} (t_1,t_2)\leqslant a \cdot\mathcal{IC}\overline H_{X,Y}(\varphi^{-1}(t_1),\varphi^{-1}(t_2)).
$$
If $\varphi$ is decreasing with $a\leqslant-\varphi'\leqslant b,~a,b>0$, then
$$
 b \cdot\mathcal{IC}H_{X,Y}(\varphi^{-1}(t_2),\varphi^{-1}(t_1))\leqslant\mathcal{IC}\overline H_{\varphi(X),\varphi(Y)}(t_1,t_2)\leqslant a \cdot\mathcal{IC}H_{X,Y}(\varphi^{-1}(t_2),\varphi^{-1}(t_1)).
$$
Moreover, Theorem \ref{t4.6} and Remark \ref{r4.2} also allow to obtain analogous results for
DCRI and DCPI with the additional assumption that $\varphi(\infty)=\infty$ and $\varphi(0)=0$, respectively.
\end{r1}
\section{Conclusions}
In recent years, there has been a great interest in the study of information measures based on
distribution functions, namely cumulative residual entropy (CRE) and cumulative past entropy (CPE).
The basic idea is to replace the density function by survival or distribution function in Shannon's
entropy. These measures possess more general properties than the Shannon entropy.
Another important generalization of Shannon entropy is the Kerridge
inaccuracy measure, which plays an important role in statistical inference, estimation and coding
theory. The concept of cumulative residual and past inaccuracy (CRI and CPI) measure has been
introduced in this paper in order to extend CRE and CPE, respectively. We studied some properties
of CRI and CPI, and their dynamic versions. Some bounds and inequalities have been obtained.
We also considered CRI and CPI for doubly truncated random variables. Several properties,
including monotonicity, and bounds have been obtained.\\
\hspace*{.2in}
The proposed measures may help information theorists and reliability analysts to study the various
characteristics of a system when it fails between two time instants. The results presented here
generalize the related existing results in context with CRE and CPE for left, right and two-sided
truncated random variables. This article is just a first step in the study of these measures;  new
properties are still under investigation.
\section*{Acknowledgements} 
We thank an anonymous referee for his/her useful comments and suggestions on the earlier version of the paper.

\end{document}